\documentclass[aps,pra,twocolumn,superscriptaddress]{revtex4-1}

\usepackage{graphicx}
\usepackage{hyperref}
\usepackage{amsmath}
\usepackage{amssymb}
\usepackage{epstopdf}
\usepackage{multirow}
\usepackage{threeparttable}
\usepackage{array}
\usepackage{booktabs}
\usepackage{xcolor}
\usepackage{mathtools}
\usepackage{ulem}

\begin{document}

\title{Wide-field mid-infrared cavity-enhanced upconversion imaging}
\author{Yue Song}
\affiliation{State Key Laboratory of Precision Spectroscopy, and Hainan Institute, East China Normal University, Shanghai 200062, China}

\author{Jianan Fang}
\email{jnfang@lps.ecnu.edu.cn}
\affiliation{State Key Laboratory of Precision Spectroscopy, and Hainan Institute, East China Normal University, Shanghai 200062, China}
\affiliation{Chongqing Key Laboratory of Precision Optics, Chongqing Institute of East China Normal University, Chongqing 401121, China}

\author{Wen Zhang}
\affiliation{State Key Laboratory of Precision Spectroscopy, and Hainan Institute, East China Normal University, Shanghai 200062, China}

\author{Yijing Li}
\affiliation{State Key Laboratory of Precision Spectroscopy, and Hainan Institute, East China Normal University, Shanghai 200062, China}

\author{Ben Sun}
\affiliation{State Key Laboratory of Precision Spectroscopy, and Hainan Institute, East China Normal University, Shanghai 200062, China}

\author{Zhiwei Jia}
\affiliation{Laboratory of Solid-State Optoelectronics Information Technology, Institute of Semiconductors, Chinese Academy of Sciences, Beijing 100083, China}

\author{Kun Huang}
\email{khuang@lps.ecnu.edu.cn}
\affiliation{State Key Laboratory of Precision Spectroscopy, and Hainan Institute, East China Normal University, Shanghai 200062, China}
\affiliation{Chongqing Key Laboratory of Precision Optics, Chongqing Institute of East China Normal University, Chongqing 401121, China}
\affiliation{Collaborative Innovation Center of Extreme Optics, Shanxi University, Taiyuan, Shanxi 030006, China}

\author{Heping Zeng}
\affiliation{State Key Laboratory of Precision Spectroscopy, and Hainan Institute, East China Normal University, Shanghai 200062, China}
\affiliation{Chongqing Key Laboratory of Precision Optics, Chongqing Institute of East China Normal University, Chongqing 401121, China}
\affiliation{Shanghai Research Center for Quantum Sciences, Shanghai 201315, China}
\affiliation{Chongqing Institute for Brain and Intelligence, Guangyang Bay Laboratory, Chongqing, 400064, China}

\begin{abstract}
Mid-infrared (MIR) spectral imaging enables precise target identification and analysis by capturing rich chemical fingerprints, which calls for high-sensitivity broadband MIR imagers at room temperature. Here, we devise and implement a continuous-wave pumping MIR upconversion imaging system based on external-cavity enhancement, which favors a large field of view, a low cavity loss, and a high spectral resolution. The involved optical cavity is constructed in an integrated fashion by utilizing one crystal facet as a cavity mirror, which allows a 43-fold power enhancement for the single-longitudinal-mode pump at 1064 nm. In combination with the chirped-poled crystal design,  high-fidelity and wide-field spectral imaging mapping is permitted to facilitate an acceptance angle up to 28.5$^\circ$ over a spectral coverage of 2.5--5 $\mu$m. Moreover, a thermal locking approach is used to stabilize the cavity at the high-power operation, eliminating active feedback and ensuring long-term stability.  A proof-of-principle demonstration is presented to showcase real-time observation of CO$_{2}$ gas injection dynamics. The implemented MIR upconversion imager features wide-field operation, high detection sensitivity, and compact footprint, which would benefit subsequent  applications including environment monitoring, gas leakage inspection, and medical diagnostics.
\end{abstract}

\maketitle

\section{Introduction}
Mid-infrared (MIR) spectrum covers vibrational-rotational transitions of molecules, thereby serving as the fingerprint spectral window for substance identification and analysis. Particularly, the chemical specificity renders MIR imaging useful in greenhouse gas monitoring and biomedical diagnostics \cite{Vodopyanov2020Book}. With the urgent demand to mitigate issues of global warming and air pollution \cite{Lacis2010Science}, there is growing interest in accurate monitoring and mitigation of greenhouse gases like CO$_{2}$ (at 4.3 $\mu$m), N$_{2}$O (at 4.5 $\mu$m), CH$_{4}$ (at 3.3 $\mu$m), as well as harmful environmental pollutants \cite{Santhanam2017earth}. Monitoring these trace gases provides critical data-driven insights for climate research, carbon emission tracking, industrial leak detection and agricultural emission analysis. Moreover, MIR spectral imaging offers a non-invasive diagnostic tool for breath analysis \cite{Vjekoslav2018Hoo}. For instance, monitoring CO (at 4.6 $\mu$m) levels in exhaled human breath can reveal smoking-related pulmonary disorders \cite{Edward2000Chest} or reflect heme oxygenase activity \cite{Stefan2013Jobr}. Similarly, HCHO (at 3.6 $\mu$m) concentrations in breath correlate with tumor metabolism \cite{Andreas2016ACSsensors}, providing a direct link to lung cancer and chronic inflammatory diseases. Typically, the aforementioned applications require gas detection sensitivities ranging from ppm to ppt levels, necessitating ultra-sensitive broadband MIR detection.

\vspace{\baselineskip}

However, MIR detectors are typically achieved using narrow-bandgap semiconductor materials (\textit{e.g.}, HgCdTe, InSb, or PbSe) \cite{Razeghi2014RPP}, which suffer from high dark current noise at room temperature. Consequently, stringent cryogenic operation is usually required to enhance the detection sensitivity. For instance, cryogenically cooled HgCdTe avalanche photodiode (APD) arrays permit MIR single-photon detection \cite{Xiaoli2019OptEng}, albeit with a limited pixel number. Recent advances in superconducting nanowire single-photon detectors (SNSPDs) have demonstrated impressively broadband responses in far-infrared wavelengths \cite{Taylor2023Optica,Pan2022OE,Chen2021SB}. Notably, the development of sensitive MIR sensors at room temperature has been fueled by emerging platforms based on low-dimensional materials \cite{Fang2020LPR, Wang2019Small, Wu2021NR, Xue2023LSA} or novel nanophotonic structures \cite{Xomalis2021Science, Chen2021Science}. In combination with high-brightness and wavelength-tunable quantum cascade lasers (QCLs), these MIR detectors enable direct acquisition of target-specific spectral image information with exceptional accuracy \cite{Haase2016JB, Shi2020NatMethods, Yeh2015AC}, although practical deployment still faces challenges in detection sensitivity, pixel scalability, and frame rates.

\vspace{\baselineskip}

Alternatively, frequency upconversion imaging serves as an indirect yet effective approach to achieve ultra-sensitive MIR detection at the single-photon level \cite{Barh2019AOP, Huang2021PR, Cai2024SA}. This method leverages nonlinear processes to transduce MIR information into the visible or near-infrared regime for accessing high-performance optical manipulation and detection \cite{Huang2022NC, Wolf2017OEA}. Recent advances in the pulse-pumped upconversion scheme favor high conversion efficiency due to the intensive peak power, which facilitates single-photon 3D imaging \cite{Fang2023LSA, Rehain2020NC}, high-speed hyperspectral imaging \cite{Fang2024NC,Zhao2023NC,Junaid2019Optica}, and edge-enhanced imaging \cite{Wang2021LPR, Zeng2023LPR}. However, such an active imaging modality typically requires tight synchronization and precise overlap to the ultrashort illumination source, which impedes efficient detection for temporally dispersed signals or randomly emitting fluorescences. Moreover, the intrinsic broadband nature of ultrashort pump pulses inevitably degrades the spectral conversion fidelity, thus limiting the resolution in spectroscopic applications.

In this context, continuous-wave (CW) pumping is proposed to realize high-resolution MIR spectral imaging \cite{Hu2012OL}. One challenge for the passive scheme is to provide high pump power, especially for the single-pass configuration \cite{Zheng2023PRA}. While photonic waveguides can significantly enhance the intensity due to the tight mode confinement and long interaction length \cite{Neely2012OL, Buchter2009OL}, the single-spatial-mode operation fundamentally precludes the wide-field imaging capability. One possible remedy is to resort to cavity enhancement techniques, where the nonlinear crystal is directly placed inside the laser resonator to implement intracavity power enhancement \cite{Witinski2009AO, Huang2019OE}. This configuration facilitates an intracavity pump power up to about 100 W, leading to demonstrations of MIR photon-counting lidar \cite{Yue2022RS, Widarsson2022AO} and single-photon spectral imaging \cite{Dam2012NP}. However, the spectral proximity between the laser pump and the upconverted signal makes it difficult to suppress background noises via spectral filtering \cite{Pedersen2019PTL}. In contrast, the so-called external-cavity enhancement scheme spatially separates the pump source from the optical cavity, favoring noise suppression and system flexibility \cite{Liu2024APN, Albota2004OL, Wolf2017OE}. Moreover, the configuration could benefit from the existing single-frequency lasers \cite{Fu2017JOSAB}, which enables high-resolution spectral measurements. Therefore, it's desirable to adopt the external-cavity scheme to realize high-efficiency and low-noise MIR upconversion imaging.

Here, we have proposed and implemented an external-cavity-enhanced MIR upconversion imaging system for the first time, which features high detection sensitivity, wide field of view, and broadband spectral coverage. The use of a chirped-poling nonlinear crystal enabled an unprecedented acceptance angle up to 28.5$^\circ$, over a wide-ranging operation wavelengths from 2.5 to 5 $\mu$m. The optical cavity is specially designed by using the crystal facet as one mirror, which favors a reduced optical loss and a more compact layout. The resulting finesse is optimized to 149, corresponding to a power enhancement factor of 43. Moreover, a thermal locking mechanism has been investigated to ensure long-term stability, eliminating the need of active feedback. Additionally, we demonstrated video-rate wide-field MIR imaging of dynamic CO$_{2}$ jets, paving the way toward spectroscopic applications.

\vspace{\baselineskip}

\section{Experimental setup}
\begin{figure*}[t!]
	\includegraphics[width=0.83 \textwidth]{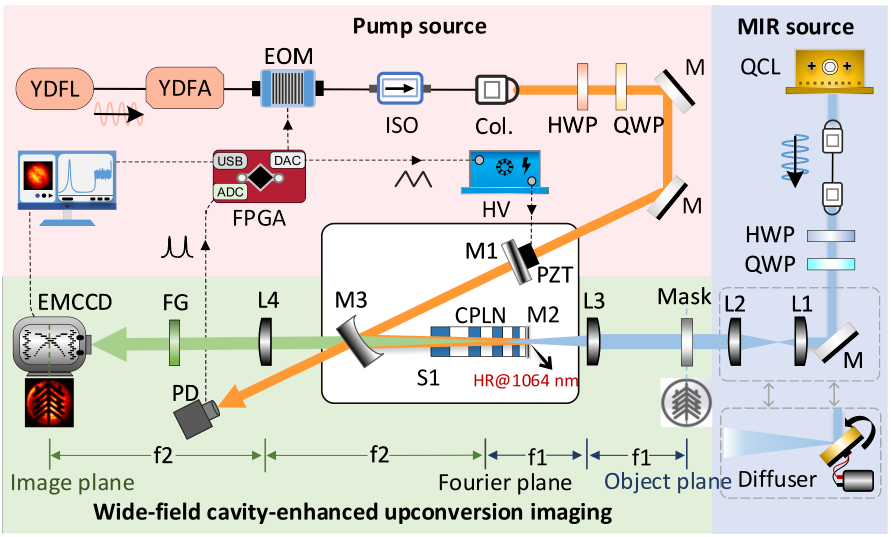}
	\caption{Experimental setup of MIR upconversion imaging based on the external-cavity pumping. The pump source from an YDFL operates at the single-longitudinal mode at 1064 nm. The QCL serves as the MIR signal at tunable wavelength around 4.3 $\mu$m. Then, the CW MIR beam is expanded with a pair of lenses before being steered into an optical cavity to implement the sum-frequency generation. The cavity is comprised of a plat mirror M1, a concave mirror M3, and a CPLN crystal. One crystal facet is coated with high reflection at 1.064 $\mu$m, thus serving the end mirror. This cavity is stabilized with a digital locking unit based on a programmed FPGA. Under the locked state, the pump power can be significantly enhanced. The MIR signal after transmitting the object is then steered into a 4f imaging system, where the crystal is placed at the Fourier plane. The upconverted image is recorded by an EMCCD after passing through a series of spectral filters. YDFL and YDFA: Yb-doped fiber laser and amplifier; EOM: electro-optical modulator; L: lens; M: mirror; Col.: collimator; ISO: isolator; PZT: piezoelectric actuator; PD: photodiode; HV: high-voltage amplifier; FG: spectral filter group.}
	\label{fig1}
\end{figure*}

Figure \ref{fig1} illustrates the experimental setup for the cavity-enhanced MIR upconversion imaging system. The pump source is from an Yb-doped fiber laser (YDFL) at 1064 nm. The YDFL is designed to stably operate at the single-frequency regime without suffering from the mode-hopping effect \cite{Fu2017JOSAB}, and its linewidth is less than 3 kHz. The YDFL delivers CW linearly-polarized light at an average power of 39 mW, which is then boosted to 1 W by using an Yb-doped fiber amplifier (YDFA). To avoid back-reflection disturbance on the YDFL, a fiber optical isolator is employed. Subsequently, the pump beam is collimated using an aspheric lens. By rotating half-wave and quarter-wave plates, the pump's polarization can be finely tuned. The output of YDFA is coupled into the optical cavity, enabling subsequent pump power enhancement. Mode matching to the external cavity is achieved by optimizing both the fiber-to-lens distance and the collimator-to-cavity spacing \cite{Liu2024APN}. The cavity finesse $\mathcal{F}$ is evaluated to be about 149 from the ratio between the free spectral range and the cavity bandwidth.

Meanwhile, the MIR source is generated by a QCL, and can be tuned from 4301 to 4310 nm by controlling the injection current. The maximum output power is 66 mW. The MIR light is collected into a single-mode fluoride fiber (P3-32F-FC-1, Thorlabs) with a coupling efficiency of 56.5\%, which allows easy light delivery and spatial mode cleaning.  Then, the MIR beam is expanded to a 1-inch diameter using a pair of CaF$_{2}$ lenses L1 and L2. Moreover, the MIR beam is modulated by a spinning diffuser before illuminating the object, to investigate the effect of spatial coherence on the imaging performance.

The MIR signal is sent into an upconversion 4f imaging system, where a chirped-poling lithium niobate (CPLN) crystal of 3$\times$2$\times$10 mm$^{3}$ (width$\times$thickness$\times$length) is placed at the Fourier plane to perform sum-frequency generation (SFG). The focal lengths of two relay lenses (L3 and L4) are 50 and 100 mm respectively. The CPLN is fabricated with linearly-ramping poling periods from 16 to 24 $\mu$m. The resulting enlarged phase-matching acceptance angle leads to a wider field of view (FOV) than that for single-poling crystals \cite{Huang2022NC}. The upconverted light at 853 nm is steered through a spectral filter group (FG) consisting of a notch filter at 1064 nm, a long-pass filter with a cut-off wavelength of 700 nm and a short-pass filter at 900 nm. The total transmission efficiency of the FG is about 90\% with a rejection ratio for the pump is about 140 dB. Finally, the upconverted image is recorded by an electron multiplying CCD (EMCCD). Notably, the implemented upconverter is intrinsically broadband with a spectral coverage of 2.5-5 $\mu$m. 

\vspace{\baselineskip}
\vspace{\baselineskip}

\begin{figure*}[t!]
	\includegraphics[width=0.75\textwidth]{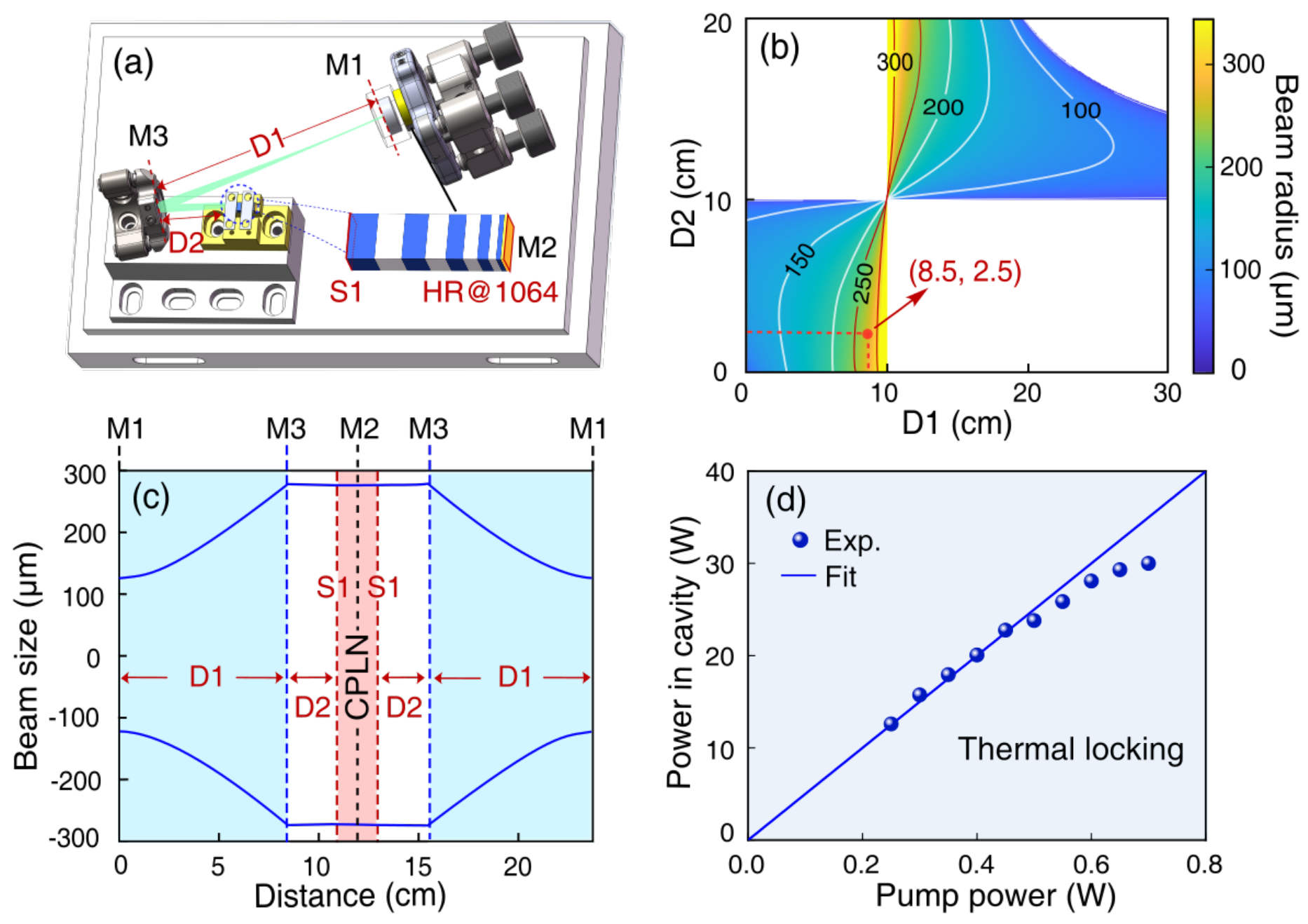}
	\caption{Optical cavity design and characterization. (a) Diagram of the semi-monolithic optical cavity. The cavity consists of a mirror M1, a concave mirror M3, and crystal end M2. The end of the crystal is coated with high reflection at 1.064 $\mu$m, which is used as a flat mirror for the cavity. (b) Beam radius within the nonlinear crystal as a function of distances of  D1 and D2. The red point denotes the beam waist used in the experiment. Note that the white area represents the unstable region for an optical cavity. (c) Evolution of the beam waist along the optical cavity, when M1 is defined as the starting position. The shaded red area denotes the nonlinear crystal. (d) The intracavity power varies with the injected pump power. The thermal locking technique is used to stabilize the optical cavity at the high-power operation.}
	\label{fig2}
\end{figure*}

\section{Results and discussion}
\subsection{Design of semi-monolithic optical cavity}
In the following, we begin with presenting the design and characterization of the optical cavity. Figure \ref{fig2}(a) gives the schematic diagram of the implemented cavity, consisting of a flat mirror (M1), a CPLN with a reflective coating on one end (M2), and a concave mirror (M3) with a curvature of 200 mm. The power reflection at 1064 nm of these three cavity mirrors are 96.87\% (M1), 99.9\% (M2) and 99.95\% (M3), respectively. Mirrors M1 and M3 are made of N-BK7. The coated crystal facet serves as one cavity mirror, enabling both a lower cavity loss and an improved integrability. Such a monolithic cavity design is a good comprise between aligning flexibility and system stability. The CPLN crystal is mounted in a brass holder and secured on an aluminum plate. These materials favor a superior thermal conductivity, thus facilitating the heat dissipation to improve the thermal equilibrium. The CPLN crystal operates at room temperature without the need of precise temperature control to satisfy the phase-matching condition. Notably, M1 is mounted with a donut-shaped piezoelectric actuator (PZT, Thorlabs PA44M3KW), which offers an injecting port for the pump. The leaking light through M3 is detected by a photodiode, serving as the monitoring port for the cavity resonance. Finally, the entire optical external cavity is integrated onto an aluminum block with geometrical dimensions of 250$\times$135$\times$105 mm$^{3}$.

\begin{figure*}[t!]
	\includegraphics[width=0.9 \textwidth]{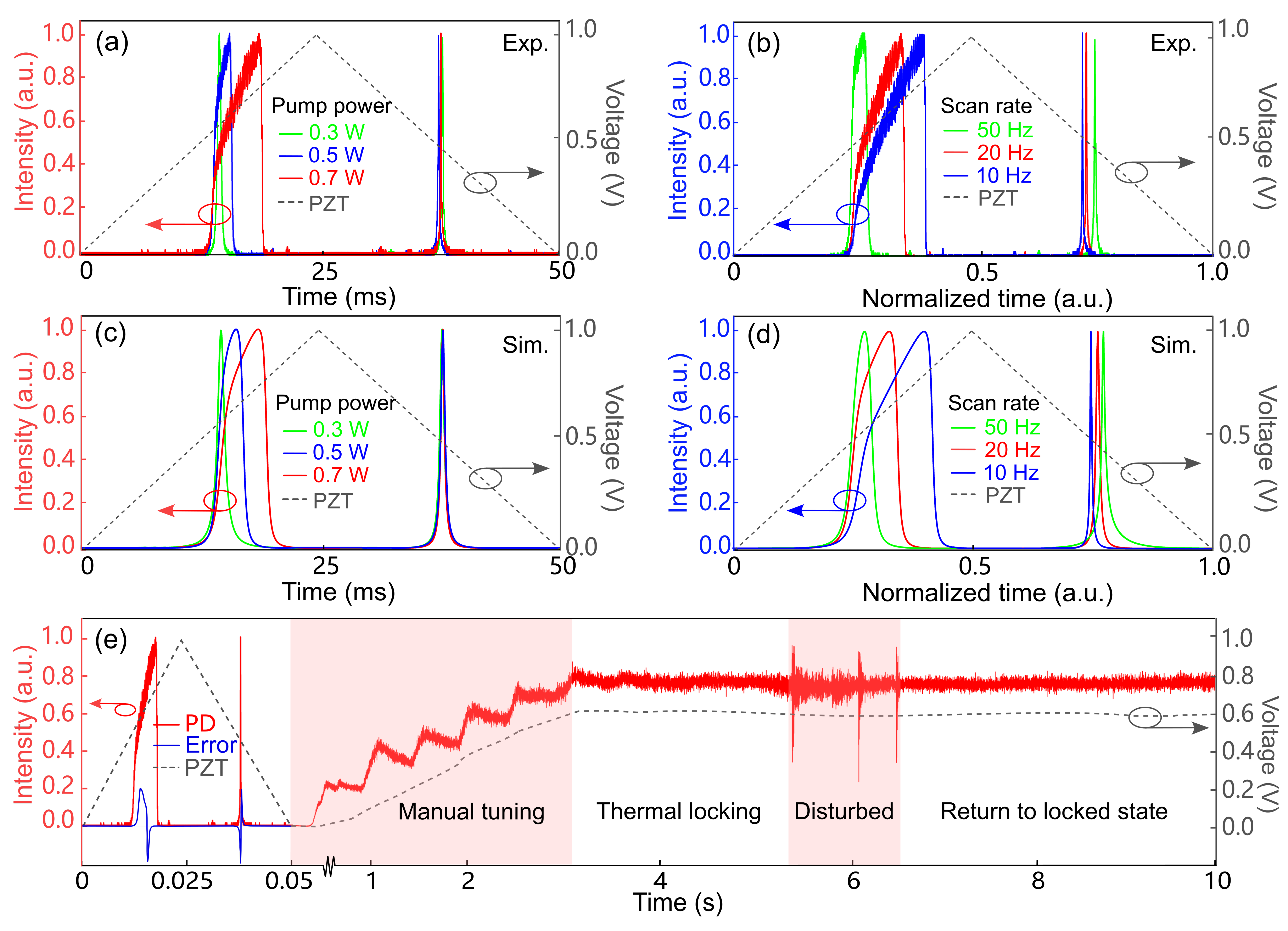}
	\caption{Thermal effect on the crystal-imbedded optical cavity.  (a, b) Experimental measurements of transmission peaks during the cavity-length scanning via the PZT in the case of various pump powers (a) and sweep rates (b). (c, d) Corresponding Numerical simulations at different pump powers (c) and sweep rates (d). (e) Process diagram of the optical locking based on the thermal effect. The locking operation initiates at 0.05 s, and stabilizes after about 2.6 s. The signal vibration is due to the external disturbance on purpose for verifying the robustness. Note that the behavior of the PDH locking is presented for the sake of direct comparison, which shows pronounced thermal effect on the error signal.}
	\label{fig3}
\end{figure*}

The cavity-mode size can be precisely controlled by tuning the mirror separations, yielding a 535-$\mu$m waist diameter at D1=8.5 cm and D2=2.5 cm as shown in Fig. \ref{fig2}(b). This optimized beam size ensures a high pump intensity and a high spatial resolution for the subsequent upconversion imaging. The resulting Rayleigh length for the Gaussian mode is sufficiently long to avoid the crystal insertion loss. Figure \ref{fig2}(c) presents the beam-size evolution along the optical cavity. For a crystal thickness of 2 mm, the beam can propagate through the nonlinear crystal without suffering from beam truncation. Thanks to the spatial mode filtering for the pump, a nearly perfect mode matching is achieved with negligible high-order peaks. The presence of only the fundamental mode facilitates the cavity locking at a stable transverse mode. To ensure a stable cavity resonance at the high-power operation, a thermal locking technique is proposed and implemented. The power in the cavity increases linearly with the power of the pump seed, as shown in Fig. \ref{fig2}(d). The power enhancement factor is deduced to be 50 from the line slope, favoring the access to high pump intensity with modest seed powers. When the cavity power exceeds 20 W, significant thermal expansion and lensing effects in the nonlinear crystal degrade spatial mode matching and introduce additional cavity losses, thereby reducing the finesse. As a result, in the presence of a pump input power of 0.7 W, the system achieves an intracavity power of 30 W, which enables a conversion efficiency of 0.1\%. Note that the achieved modest efficiency is a common comprise for broadband upconversion imagers \cite{Fang2024NC, Mrejen2020LPR}, and favors a much lower injection pump power comparing to single-pass configuration \cite{Zheng2023PRA}.

\subsection{Thermal locking for cavity stabilization}
Next, we further analyze the enhancement performance under thermal effects, which are manifested by shifts in transmission peaks when applying a high-voltage sawtooth waveform to the PZT for cavity-length sweeping. Figure \ref{fig3}(a) presents the behavior in the case of various injecting pump power. The scanning rate is fixed at 20 Hz. As the pump power increases from 0.3 to 0.7 W, the asymmetry in the transmission peaks corresponding to the up- and down-scanning directions becomes more pronounced. Specifically, the up-scanning peak broadens, while the down-scanning peak becomes narrower. The underlying mechanism lies in the power-dependence length elongation due to the thermal expansion. When a higher voltage is applied to the PZT, the cavity length decreases (see Fig. \ref{fig2}(a)). As the cavity approaches the resonance position, the intra-cavity power increases significantly. In response, the effective optical-path length of the crystal extends to compensate for the cavity-length reduction induced by the PZT. As a result, the resonating position is delayed relative to the ``normal" one. Meanwhile, the peak looks spread during its positive slope, corresponding to a ``slower scan". In the down-scanning case, the cavity length extends as the voltage decreases. Consequently, the thermal-induced elongation accelerates the arrival on the resonance, thus resulting in a ``faster scan" with a narrower peak width. Such a phenomenon would be more evident for a slower scanning speed due to the longer time to heat the crystal. Figure \ref{fig3}(b) presents the experimental observations at various scanning rates for a fixed injection power of 0.7 W. At a slower scan rate of 10 Hz, sufficient time is available for thermal accumulation, leading to a pronounced extension of the crystal's optical path. This results in a significant delay of the cavity resonance during the up-scan. In contrast, at a higher scan rate of 50 Hz, the thermal effect is mitigated due to the relatively long relaxation time, thereby reducing the distortion of the transmission peaks.

To elucidate this phenomenon, a numerical simulation is performed to reveal the cavity dynamics in the presence of the thermal effect. Assuming the Fabry-P\'{e}rot cavity is impedance matched and lossless, the transmission power $P_{\text{trans}}$ is given by \cite{Carmon2004OE}:
\begin{align}
 \label{eq1}
 \frac{P_{\text{trans}}}{P_{0}} =\frac{1}{1+\mathcal{F}\cdot \sin^{2}(kd)},
\end{align}
where $\mathcal{F}$ is the cavity finesse, $P_{0}$ denotes the optical cavity input power, $k$ is the wave vector and $d$ is the optical length of the cavity. When accounting for thermal-expansion-induced path length variations, the phase $kd$ depends on the temperature of the cavity. It can be expressed as $kd = w_{0}(d_{0}+vt)(1+\beta \Delta T(t))/c$, where $w_{0}$ is the frequency of the light, $c$ is the speed of light, $d_{0}$ denotes the initial cavity length at ambient temperature, $v = \text{d}d/\text{d}t$ represents the cavity length sweep speed, $\beta$ is a coefficient accounting for both material thermal expansion and the change in refractive index with temperature. We assume the crystal's temperature response as a first-order impulse function $h(t)$ with characteristic time constant $\tau_{0}$, $i.e.$, $h(t) = (1/\tau_{0})\text{exp}(-t/\tau_{0})$ when $t > 0$. Then $\Delta T(t) = \alpha P_{\text{trans}}(t)\ast h(t)$, $P_{\text{trans}}(t)\ast h(t)$ represents the convolution of these two terms, and $\alpha$ is a positive constant. The simulation results for the transmission traces are plotted in Fig \ref{fig3}(c,d) under different incident pump powers and scanning rates, which show good agreement with the measured responses.

Intriguingly, a passive thermal locking of the external optical cavity can be realized by leveraging the thermal effect, where the resulting broadening of the resonance peak effectively extends the range of cavity lengths that can sustain resonance. Specifically, thermal self-locking is accomplished by up-scanning the PZT and halting the scan just before the maximum intracavity power is reached, thus avoiding transition into the thermally unstable regime. Although the thermal locking confines the cavity operation to approximately 85\% of the theoretical maximum power, it eliminates the need for complex active-feedback stabilization systems compared to previous works \cite{Liu2024APN}. Figure \ref{fig3}(e) shows the intracavity power dynamics under the thermal locking. The input pump power is set to 0.7 W, with a PZT scanning rate of 20 Hz. Asymmetry in the error signal in the conventional Pound-Drever-Hall (PDH) technique can be observed, thus affecting the ability to stabilize the cavity precisely at the zero-crossing of the error signal. By manually tuning the high-voltage driver to initiate thermal locking, we can achieve a stable enhancement of the intracavity power. In the presence of repetitive mechanical shocks, the cavity returns to the locked state within a short time about 0.1 s, demonstrating strong resilience against disturbances and robust self-recovery. Moreover, the chirped poling period design of the crystal enables a wide phase-matching bandwidth, which allows a larger tolerance for the heating-induced temperature variation. As a result, the conversion efficiency would be insensitive to thermal fluctuations while maintaining the locking stability of the optical cavity. Finally, long-term operational stability is achieved with optical power fluctuations of approximately 2.5\%.

\subsection{Wide-field cavity-enhanced imaging}
\begin{figure}[t!]
	\includegraphics[width=0.95 \columnwidth]{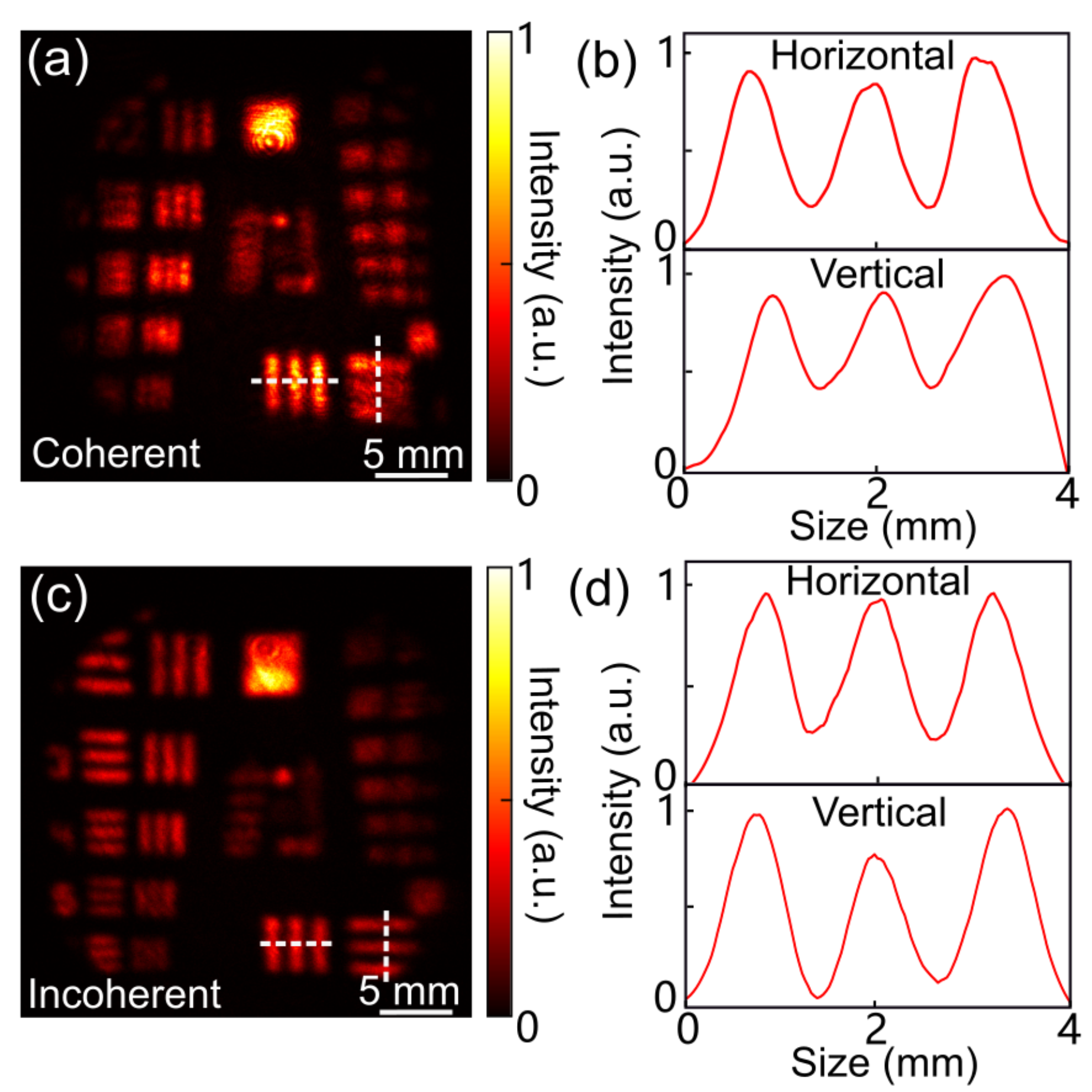}
	\caption{Performance characterization of the MIR upconversion imaging under coherent and incoherent illumination. (a) Coherent MIR upconversion images for the USAF resolution target. (b) Representative cross-section traces are given for the line pairs of the first element in the zeroth group under the coherent illumination. (c) Incoherent MIR upconversion images. (d) Corresponding cross sections, which show an enhanced contrast. Note that all the images are acquired by an EMCCD at an exposure time of 80 ms.}
	\label{fig4}
\end{figure}

Now, we turn to investigate the performance of the MIR upconversion imager under both coherent and incoherent illumination conditions. A coherent MIR beam from the QCL is expanded to a 1-inch diameter and directed onto a USAF 1951 resolution target. The resulting upconverted image is shown in Fig. \ref{fig4}(a). Due to the aperture size of the input lens L3 in the 4f system, the field of view is constrained to approximately 1 inch in diameter. The smallest resolvable bars, indicated by dashed lines, correspond to element 1 in group 0, with a spatial linewidth of 500 $\mu$m. The corresponding intensity cross sections are displayed in Fig. \ref{fig4}(b), showing fringe contrasts of approximately 58\% and 48\% along the horizontal and vertical directions, respectively. The theoretical resolution limit of the system is given by $R = 4\lambda_s f_1 / \pi D_p$, where $\lambda_s$ is the wavelength of the MIR signal, $f_1$ = 50  mm is the focal length of lens L3, and $D_p$ is the pump beam waist within the nonlinear crystal. Increasing the pump beam waist enables the conversion of higher spatial-frequency components, thereby improving spatial resolution. However, a larger beam waist reduces the pump intensity, which in turn decreases the overall upconversion efficiency. To balance spatial resolution and detection sensitivity, the pump beam waist is optimized to approximately 535 $\mu$m. The corresponding theoretical resolution of 513 $\mu$m agrees well with the experimentally observed value.

For incoherent signals, the upconverted image corresponds to the convolution of the real-valued intensity distribution of the rescaled object with the intensity profile of the pump's Fourier transform. Since this convolution is performed on intensity rather than field amplitude, the spatial resolution is theoretically improved by a factor of $\sqrt{2}$ compared to that in the coherent imaging regime. Experimentally, the MIR beam is modulated using a spinning diffuser prior to illuminating the object. This approach effectively suppresses speckle artifacts typically observed in coherent imaging due to the high spatial coherence of the laser source. To enable a direct comparison with the coherent case, the same resolution target is imaged under incoherent illumination. The resulting upconverted image is shown in Fig. \ref{fig4}(c), revealing improved image uniformity and enhanced detail visibility. Cross-sectional intensity profiles are presented in Fig. \ref{fig4}(d), where the resolved bar width is measured to be 353 $\mu$m, close to the theoretical prediction.

\begin{figure*}[t!]
	\includegraphics[width=0.95 \textwidth]{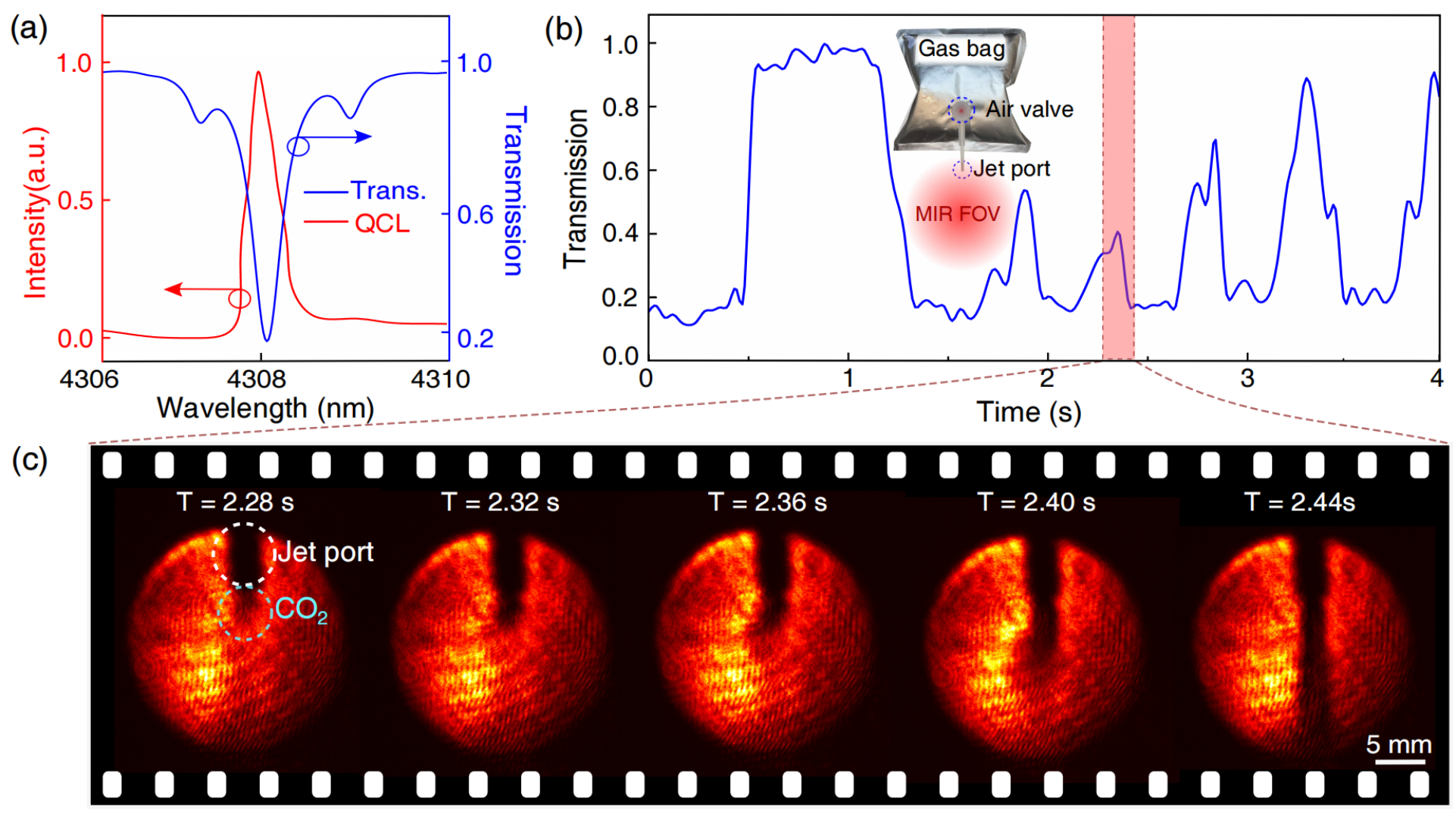}
	\caption{Real-time MIR spectral imaging for gas flow monitoring. (a) Transmission spectrum of carbon dioxide at the standard atmospheric pressure, which is obtained from the Hitran database. The optical spectrum of the used QCL light source centers at 4308 nm, coinciding with the gas absorption peak. (b) Real-time monitoring of the CO$_{2}$ concentration within the imaging FOV. The CO$_{2}$ gas is delivered from a pre-filled reservoir bag and injected into the imaging area through a nozzle, as illustrated in inset. By repeatedly compressing the gas bag, gas concentration varies in time as expected. (c) Spectral images during the period of 2.28 to 2.44 s. The MIR illumination power is about 10 $\mu$W/mm$^{2}$. The exposure time of the camera is set to 40 ms, corresponding to a frame rate of 25 fps. See Supplemental Video 1 for the recorded dynamics.}
	\label{fig5}
\end{figure*}

\subsection{Real-time monitoring of gas flow dynamics}
Finally, we demonstrate the real-time CO$_2$ gas monitoring capability of the MIR imaging system. As shown in Fig. \ref{fig5}(a), CO$_2$ exhibits a strong absorption peak at 4308 nm. The narrow linewidth of the QCL enables precise spectral alignment with this feature, with the measured spectral width limited by the spectrometer's resolution (approximately 0.1 nm). The experimental setup is illustrated in Fig. \ref{fig5}(b). The CO$_2$ is introduced into the imaging field of view under ambient conditions, with standard atmospheric pressure (1 atm) and room temperature (300 K). A sealed aluminum-foil gas bag filled with CO$_2$ serves as the source, connected to a fine-aperture nozzle for controlled injection into the FOV. The flow rate is precisely regulated using a needle valve to ensure consistent delivery and stable diffusion dynamics.

To better illustrate the CO$_2$ absorption process, we select a representative point along the gas flow path to monitor its MIR transmittance, as shown by the dynamic curve in Fig. \ref{fig5}(b). The transmittance exhibits temporal variations over a 4-second measurement window (see Supplementary Video 1 for the corresponding real-time imaging sequence). At \(T = 0.92\) s, when the valve is closed, the CO$_2$ concentration within the FOV matches ambient levels, and the corresponding image intensity serves as the baseline reference for normalized transmittance. Upon gas release, the transmittance drops rapidly as CO$_2$ fills the monitored region. Figure \ref{fig5}(c) captures a full cycle of the CO$_2$ diffusion process between 2.28 and 2.44 s. The EMCCD camera operates at 25 frames per second (with a 40 ms exposure time per frame and a resolution of 1024$\times$1024 pixels). Notably, the system's high sensitivity permits even faster MIR imaging when equipped with high-speed CMOS detectors. The demonstrated combination of wide field of view and high detection sensitivity underscores the system's potential for diverse applications, including environmental pollution monitoring, trace gas spectroscopy, and long-range target detection.

\section{Conclusion}
In conclusion, we have demonstrated an external-cavity-enhanced MIR upconversion imaging system, which features a wide field of view, a broad spectral coverage, and room-temperature operation. By applying anti-reflection coatings to the facets of the CPLN crystal in place of conventional cavity mirrors, we significantly reduced intracavity losses and improved the pump enhancement factor, resulting in a highly integrated cavity design. The decoupled configuration of the pump source and enhancement cavity enables the use of compact, single-frequency fiber lasers and amplifiers, thereby avoiding the spectral engineering challenges associated with solid-state lasers. This narrowband pumping scheme is particularly well-suited for high-fidelity spectral mapping, making it highly attractive for precision molecular spectroscopy. Furthermore, the adoption of a thermal locking technique eliminates the need for complex active feedback systems, while simultaneously increasing the tolerance to cavity-length variations and reducing the overall system footprint, hence beneficial for portable applications. In addition, the system facilitates real-time gas flow monitoring, showcasing its robustness and potential for precision spectroscopic imaging. By integrating either a rapidly tunable broadband QCL/OPO source \cite{Junaid2019Optica} or a supercontinuum illuminator combined with fast spectral filtering of upconverted signals \cite{Zhao2023NC}, the system is well-positioned for deployment in MIR hyperspectral imaging applications.

To go beyond the achieved performance, several avenues merit future investigation. First, the current spatial resolution is intrinsically limited by the Gaussian profile of the pump beam, which restricts the spatial frequency bandwidth in the Fourier plane. One promising approach is to engineer an optical cavity that supports elliptical pump resonances, thereby maximizing the utilization of the nonlinear crystal's cross-sectional area and enabling improved high-frequency component conversion through nonlinear Fourier ptychographic imaging \cite{TZheng2024Optica}. Second, by applying high-reflectivity pump coatings to both end faces of the nonlinear crystal, while introducing controlled curvature on one facet, the crystal can function as a plano-concave monolithic optical cavity. This architecture eliminates the need for external mirrors, leading to a compact, highly integrated design with reduced optical losses, making it well-suited for broader deployment scenarios. Additionally, the phase-matching condition in the nonlinear 4f imaging system induces wavelength-dependent radial shifts during the upconversion process, resulting in spatially dispersed broadband images \cite{Fang2024NC}. This inherent dispersion provides a practical route toward snapshot MIR hyperspectral imaging.
 
\vspace{8pt}
\noindent  {\fontfamily{phv}\selectfont 
\normalsize \textbf{Acknowledgments.} 
}
\noindent This work was funded by Shanghai Pilot Program for Basic Research (TQ20220104); National Natural Science Foundation of China (62175064, 62235019, 62035005); Innovation Program for Quantum Science and Technology (2023ZD0301000); Shanghai Municipal Science and Technology Major Project (2019SHZDZX01); Natural Science Foundation of Chongqing (CSTB2023NSCQ-JQX0011, CSTB2022TIAD-DEX0036);  Fundamental Research Funds for the Central Universities; China Postdoctoral Science Foundation (2024M760918).

\vspace{8pt}
\noindent  {\fontfamily{phv}\selectfont 
\normalsize \textbf{Disclosures.} 
}
\noindent The authors declare no competing interests.

\vspace{8pt}
\noindent  {\fontfamily{phv}\selectfont 
\normalsize \textbf{Data availability.} 
}
\noindent The data that support the findings of this study are available from the corresponding author upon reasonable request.

\end{document}